\documentclass[preprint,superscriptaddress,nofootinbib,pra,aps]{revtex4-1}

\usepackage{amsmath,amssymb,graphicx,epstopdf,mathtools,color,url,siunitx,hyperref}
\usepackage{soul}
\usepackage{physics}

\makeatletter 
\renewcommand{\pdv}[2]{\begingroup 
  \@tempswafalse\toks@={}\count@=\z@ 
  \@for\next:=#2\do 
    {\expandafter\check@var\next\@nil
     \advance\count@\der@exp 
     \if@tempswa 
       \toks@=\expandafter{\the\toks@\,}% 
     \else 
       \@tempswatrue 
     \fi 
     \toks@=\expandafter{\the\expandafter\toks@\expandafter\partial\der@var}}% 
  \frac{\partial\ifnum\count@=\@ne\else^{\number\count@}\fi#1}{\the\toks@}% 
  \endgroup} 
\def\check@var{\@ifstar{\mult@var}{\one@var}} 
\def\mult@var#1#2\@nil{\def\der@var{#2^{#1}}\def\der@exp{#1}} 
\def\one@var#1\@nil{\def\der@var{#1}\chardef\der@exp\@ne} 
\makeatother 

\allowdisplaybreaks

\begin{document}

\title{Cost of remembering a bit of information}
\author{D. Chiuchi\`u}
\thanks{Presently at Okinawa Institute for Science and Technology}
\affiliation{NiPS Lab, Universit\`a degli studi di Perugia, Dipartimento di Fisica e Geologia}
\author{M. L\'opez-Su\'arez}
\affiliation{NiPS Lab, Universit\`a degli studi di Perugia, Dipartimento di Fisica e Geologia}
\author{I. Neri}
\email{igor.neri@nipslab.org}
\affiliation{NiPS Lab, Universit\`a degli studi di Perugia, Dipartimento di Fisica e Geologia}
\affiliation{INFN, sezione di Perugia}
\author{M. C. Diamantini}
\affiliation{NiPS Lab, Universit\`a degli studi di Perugia, Dipartimento di Fisica e Geologia}
\affiliation{INFN, sezione di Perugia}
\author{L. Gammaitoni}
\affiliation{NiPS Lab, Universit\`a degli studi di Perugia, Dipartimento di Fisica e Geologia}

\begin{abstract}
In 1961, Rolf Landauer pointed out that resetting a binary memory requires a minimum energy of $k_BT \ln(2)$. However, once written, any memory is doomed to loose its content if no action is taken. To avoid memory losses, a refresh procedure is periodically performed. In this paper we present a theoretical model and an experiment on a micro-electro-mechanical system to evaluate the minimum energy required to preserve one bit of information over time. Two main conclusions are drawn: i) in principle the energetic cost to preserve information for a fixed time duration with a given error probability can be arbitrarily reduced if the refresh procedure is performed often enough; ii) the Heisenberg uncertainty principle sets an upper bound on the memory lifetime.
\end{abstract}

\maketitle

The act of remembering is of fundamental importance in human life. 
Not only manmade objects, monuments and landscapes require maintenance to counterbalance their deterioration, but also biological systems are subjected to the never-ending task of preserving shapes and functionalities by fighting the universal tendency of entropy to increase. As a consequence, the study of fundamental physical limits in memory devices \cite{landauer} has received considerable attention in different contexts in these years. Examples are in communication-theoretic paradigms \cite{mancini}, proteins functionality \cite{horowitzengland}, biological noisy neural networks \cite{burak,diamantini}, future technologies \cite{markov,horowitzjacobs} and in the presence of limited knowledge \cite{lloyd,gamma}. However the fundamental energetic cost to preserve the state of a memory has received little attention so far. In this work we investigate theoretically and experimentally the minimum energy cost required to preserve classical information stored in digital devices for a given time and with a given probability of failure.

\begin{figure}
\includegraphics[width=\linewidth]{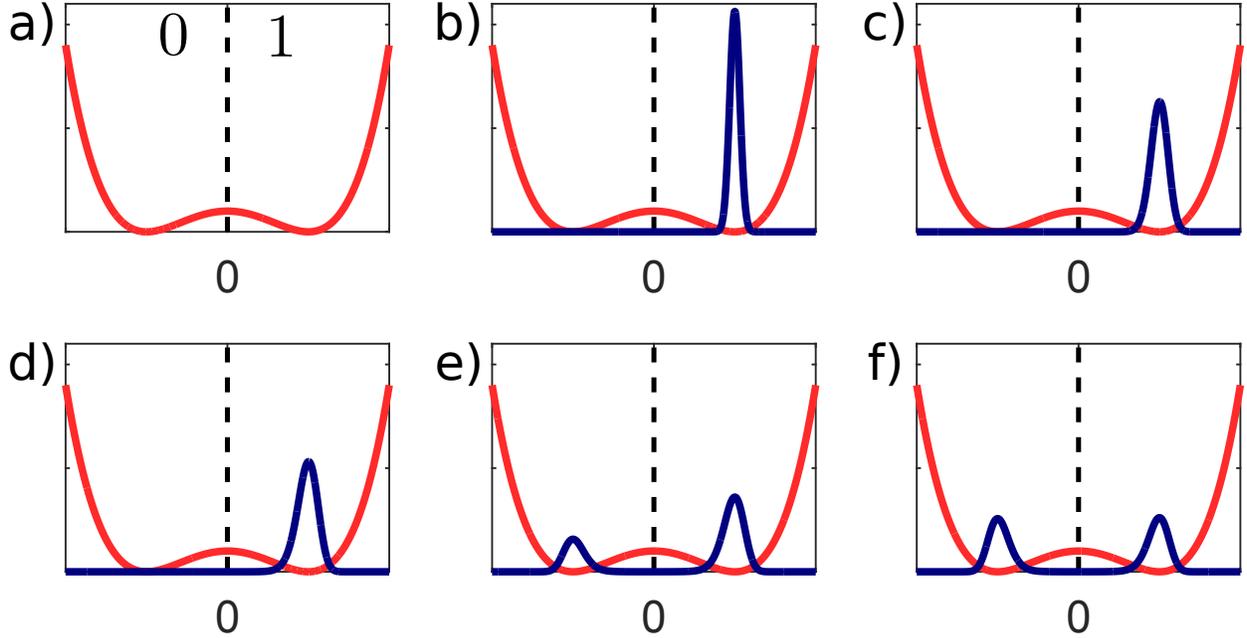}
\caption{A generic binary memory is represented here in terms of the stochastic dynamics of a variable $x$ subjected to a bistable potential (a). Panels (b) to (f) show the memory-loss mechanism when the bit 1 is initially stored. Blue (dark gray) curves give a qualitative time evolution of $p(x,t)$ as the relaxation to equilibrium process takes place.}\label{fig:figure_1}
\end{figure}

To this end we recollect that information is  usually stored  in digital devices through binary numbers (0 and 1). As a consequence, it is customary to represent a memory as a two-state physical system with an observable $x$ and a bistable potential energy landscape (Fig. \ref{fig:figure_1}.a)\cite{landauer,berut,chiuchiudiamantini,neri,lopez}. The energy barrier allows to define the two logic states, e.g. $x<0$, representing bit 0 and $x>0$, representing bit 1. Moreover, the barrier allows to statistically confine $x$ for a given time within one of the two wells (Fig. \ref{fig:figure_1}.b), hence ensuring that one given bit is stored.
This confined state is a non-equilibrium condition that evolves, within the system relaxation time $\tau_k$, to thermal equilibrium (Fig. \ref{fig:figure_1}.f). This process is described via the time evolution of the probability density function $p(x,t)$ as follows. Let us assume we have a memory where the bit 1 is stored. The initial probability density $p(x,0)$ shows a sharp peak centred in the right well (Fig. \ref{fig:figure_1}.b). According to the dynamic of the system, $p(x,t)$ will first relax inside the right well and then it will diffuse into the left well, thus developing a second peak (Fig. \ref{fig:figure_1}.b to \ref{fig:figure_1}.f). At any given time $t$, the probability that the system encodes the wrong logic state is represented by $P_0(t)=\int_{-\infty}^0 p(x,t)\mathrm{d}x$. Clearly $P_0$ increases with time and reaches the thermal equilibrium condition $P_0=0.5$ when the memory is statistically lost (Fig. \ref{fig:figure_1}.f).

To fight this natural deterioration of the bit, it is customary to perform a cyclic operation called \emph{refresh}. This procedure consists in reading and then writing back the content of the memory, and it is periodically executed at intervals $t_R$\cite{laplante2005comprehensive,bruce}. The refresh operation restores a non-equilibrium condition by shrinking the width of each peak of $p(x,t)$. Note that, during this refresh operation no error correction is performed as the overall purpose is merely to fight the diffusive process leading to thermal equilibrium.

Based on this procedure, we can define the memory loss probability $P_E$ at time $\overline{t}$, i.e. after $N=\overline{t}/t_R$ cycles, as:
\begin{equation}\label{eq:equation1}
P_E =1-\Big[1-P_0 \left(t_R \right)\Big]^{\frac{\overline{t}}{t_R}}
\end{equation}
It indicates the probability to find the wrong value of the bit when the memory is interrogated at any time during the interval $\left[0-\overline{t}\right]$ since its first writing, with a refresh interval $t_R$.
In any practical application it is interesting to \emph{a priori} set both $P_E$ and $\overline{t}$, and then deduce the optimal $t_R$ to meet these targets. Assumed that the refresh operation has an energetic cost $Q$, what we want to address here is the fundamental minimum energetic cost $Q_m$ to preserve a given bit for a time $\overline{t}$, with a probability of failure not larger than $P_E$, while executing the refresh procedure with periodicity $t_R$.
To this end we proceed as follows: we first investigate the maximum value of $t_R$ for a given set of $P_E$ and $\overline{t}$; secondly we perform an experiment to measure the minimum energetic cost for a single refresh operation; finally we estimate the physical fundamental limits associated with the overall procedure.

We start with the study of the maximum allowed value for $t_R$.
Let us assume that the dynamics of the memory is characterised by a bistable Duffing potential: 
\begin{equation}\label{eq:duffing}
U(x)=4\left(-\frac{x^2}{2}+\frac{x^4}{4}\right)
\end{equation}
The probability density function $p(x,t)$ thus evolves according to the following dimensionless Fokker-Plank equation \cite{risken,gardiner}:
\begin{equation}\label{eq:FP}
\pdv{}{t}\ p(x,t) =\pdv{}{x}\left(\pdv{U}{x}\ p(x,t) \right)+T\pdv{^2}{x^2}\ p(x,t),
\end{equation}
where $T$ is the temperature of the thermal bath. 
Solving numerically eq.\eqref{eq:FP} and using eq.\eqref{eq:equation1}, we obtain the maximum refreshing interval $t_R$ that satisfies the \emph{a priori} requirements for $\overline{t}$ and $P_E$ (see Appendix \ref{a:A}). Fig. \ref{fig:figure_2}  shows the results of this study. We can see that large time $\overline{t}$ and small probability of error $P_E$ yield short refresh time $t_R$, as expected. 
\begin{figure}
\includegraphics[width=\linewidth]{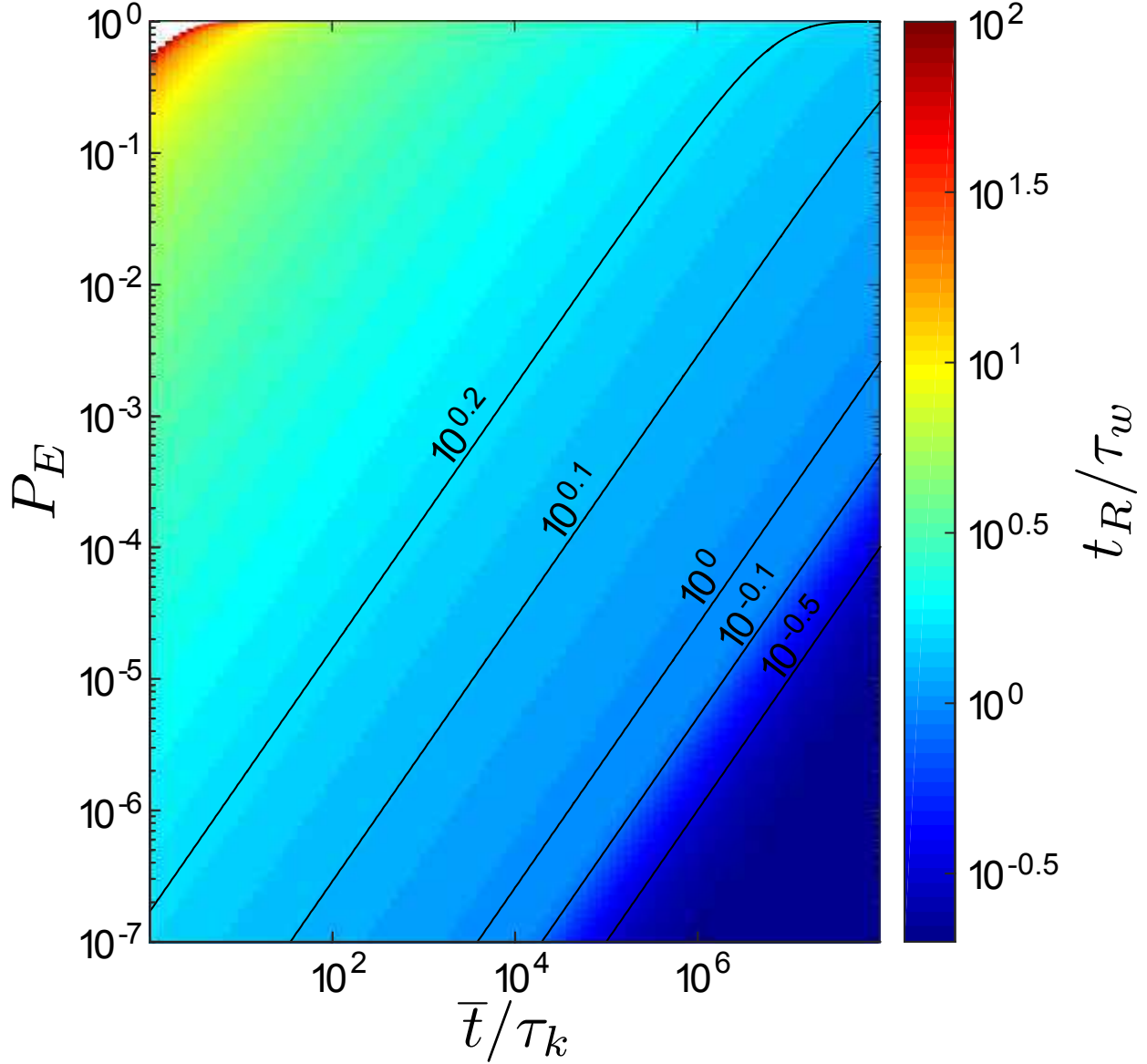}
\caption{Plot of $t_R$ as function of $\overline{t}$ and $P_E$ for a memory modeled with a bistable Duffing potentials. Here $t_R$ is given as a multiple of $\tau_w$, i.e. the relaxation time of the harmonic approximation within one well.}\label{fig:figure_2}
\end{figure}

We now proceed to the second step of our program aimed at determining the minimum energetic cost for a single refresh operation. Within the formalism defined above, the refresh operation consist in shrinking $p(x,t_R)$ inside one of the wells of $U(x)$. Thus, the energetic cost becomes a function of $t_R$ identified above.
If we assume that $t_R\ll\tau_k $, the system dynamics is practically confined within one well. Here it can be approximately described by the dynamics of an harmonic oscillator, characterised by a Gaussian probability density function \cite{chiuchiu}.
 
\begin{figure}
\includegraphics[width=\linewidth]{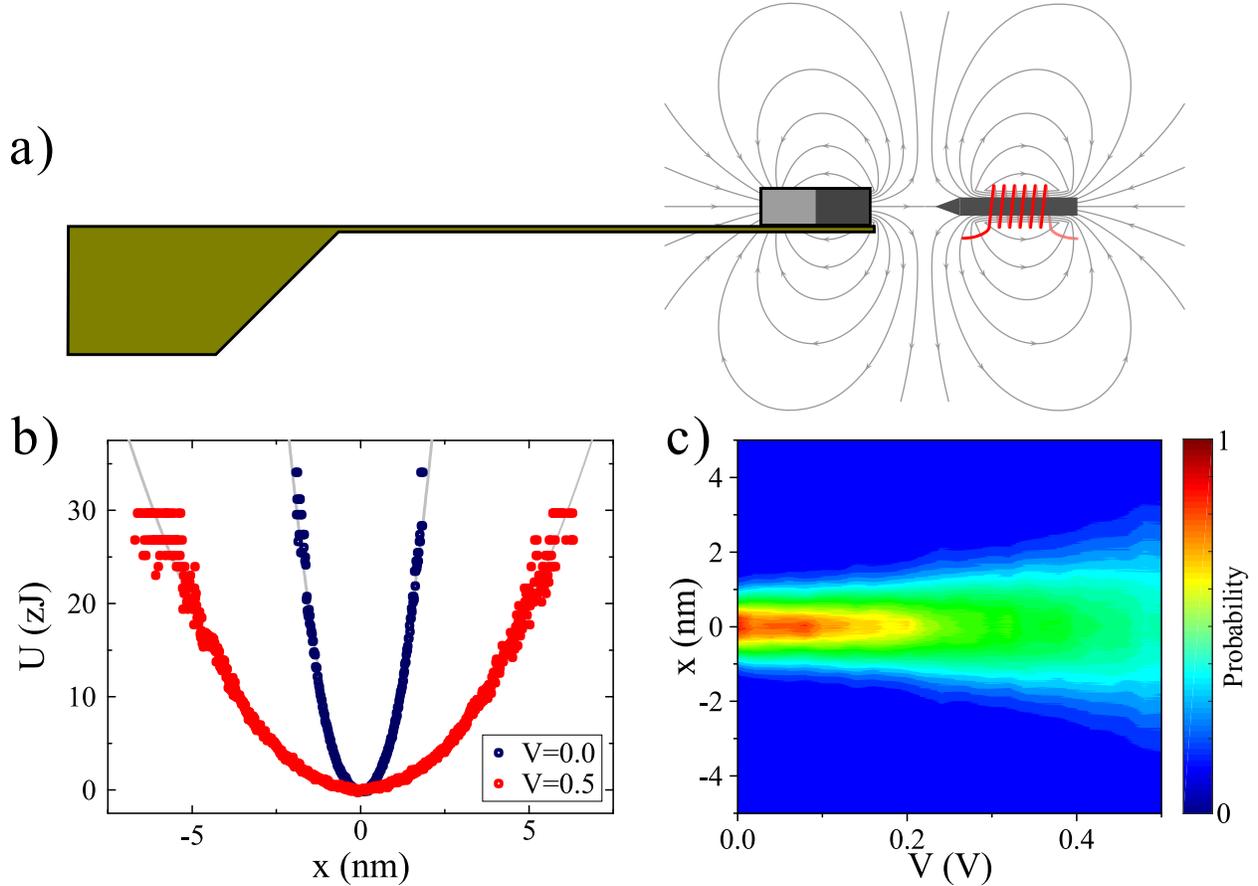}
\caption{Experimental setup. (a) Lateral view: a magnet on the cantilever tip and an electromagnetic coil are used to change the effective stiffness of the cantilever. (b) Potential energy of the cantilever tip, for two different voltages at the coil, reconstructed from the equilibrium probability density function. Solid gray lines represent the fitted harmonic potential. (c) Equilibrium probability density function of the cantilever tip position as a function of the voltage applied at the coil. The greater is the voltage, the greater is the repulsive force, resulting in a flattening of the potential and in a broadening of the equilibrium probability density function.}\label{fig:figure_3}
\end{figure}

To estimate the energy cost associated with a real refresh procedure we decided to perform an experiment employing a micro electro-mechanical oscillator composed by a $\SI{200}{\micro\metre}$ long V-shaped structure with a nominal stiffness $k=\SI{0.08}{\newton\per\metre}$, and a resonance frequency of $\SI{17}{\kilo\hertz}$. A tiny NdFeB (neodymium) magnet is attached to the cantilever tip with bi-component epoxy resin reducing its resonance frequency to $\SI{5.3}{\kilo\hertz}$. An external electromagnet  is placed in front of the cantilever as depicted in Fig. \ref{fig:figure_3}.a. 
The deflection of the cantilever, $x$, is measured with an AFM-like optical lever: a laser beam is focused on the cantilever tip with an optical lens (focal length $f=50 mm$), and a small bend of the cantilever provokes the deflection of a laser beam that can be detected with a two quadrants photo detector. For small cantilever deflections the response of the photo detector remains linear, thus $x=r_x \Delta V_{PD}$, where $\Delta V_{PD}$ is the voltage difference generated by the two quadrants of the photo detector, and $r_x$ is a calibration factor obtained through the frequency response of the system under the action of thermal fluctuations. In the small oscillation approximation the system dynamics can be modelled as a single degree-of-freedom subjected to a harmonic potential due to two forces: the cantilever restoring force and the magnetic force between the NdFeB magnet and the electromagnet. The measurement has been performed in vacuum, at pressure of $\SI{1e-3}{\milli\bar}$. In this condition the quality factor of the system is $Q_f=300$, resulting in a relaxation time $t_{Relax}=\SI{20}{\milli\second}$. The experiment is conducted at room temperature and the system is subjected to thermal fluctuations and frictional forces as well. The magnetic force can be altered over time by varying the voltage on the electromagnet. In our experiment the voltage applied to the coil results in a repulsive force with the effect of softening the potential energy of the system. The protocol used to perform the refresh operation is the following: at time $t=0$ the voltage is linearly changed from the initial value $V=0.5V$ to $V=0V$.  During this operation the effective harmonic potential changes from the one represented by the red (light gray) dots in Fig. \ref{fig:figure_3}.b to the blue (dark gray) ones. 
The equilibrium probability density function of the tip position changes accordingly as depicted in Fig. \ref{fig:figure_3}.c, from right to left.
The entire procedure takes a time $t_p$, after which the voltage on the coil is suddenly changed back to $V=0.5V$ and kept in this condition for a time $t_R$. 

The total work $W$ performed by the external force on the memory system during the refresh operation can be estimated as\cite{sekimoto,seifert}:
\begin{equation}
W =\left\langle \int_0^{\tau_p} \frac{\partial H(x,V)}{\partial V} \dot{V}\, dt\right\rangle 
\end{equation}
were $H(x, V)$ is the total energy of the system, $x(t)$ the measured trajectory of the cantilever tip, $V(t)$ is the voltage applied on the electromagnet, and $\langle \cdot \rangle$ denotes the average over an ensemble of realizations. In particular we used here $\sim 500$ experimental trajectories for each selected time protocol $\tau_p$ under study.
Since there is no variation on the internal energy of the system, the energetic cost $Q$ of a refresh operation coincides with the work performed on the system ($Q=W$). This quantity has to be compared with the thermodynamic minimum $-T\Delta S$ where (see Appendix \ref{a:B})
\begin{equation}\label{eq:equation3}
\Delta S= k_B \ln\left(\frac{\sigma_i}{\sigma_f} \right)
\end{equation}
is the entropy change associated with the refresh operation, $\sigma_i$ is the target standard deviation of the Gaussian peak to be achieved with the refresh and $\sigma_f$ is the standard deviation of the Gaussian peak before the refresh. While $\sigma_i$ can be arbitrary chosen, $\sigma_f$ depends on $t_R$ as (see Appendix \ref{a:B})
\begin{equation}\label{eq:equation2}
\sigma_f=\sqrt{\sigma_w^2+\exp\left(-\tfrac{t_R}{\tau_w}\right) (\sigma_i^2-\sigma_w^2) }
\end{equation}
where $\sigma_w$ is the equilibrium standard deviation of the harmonic oscillator and $\tau_w$ is the relaxation time of the harmonic oscillator.

\begin{figure}
\includegraphics[width=\linewidth]{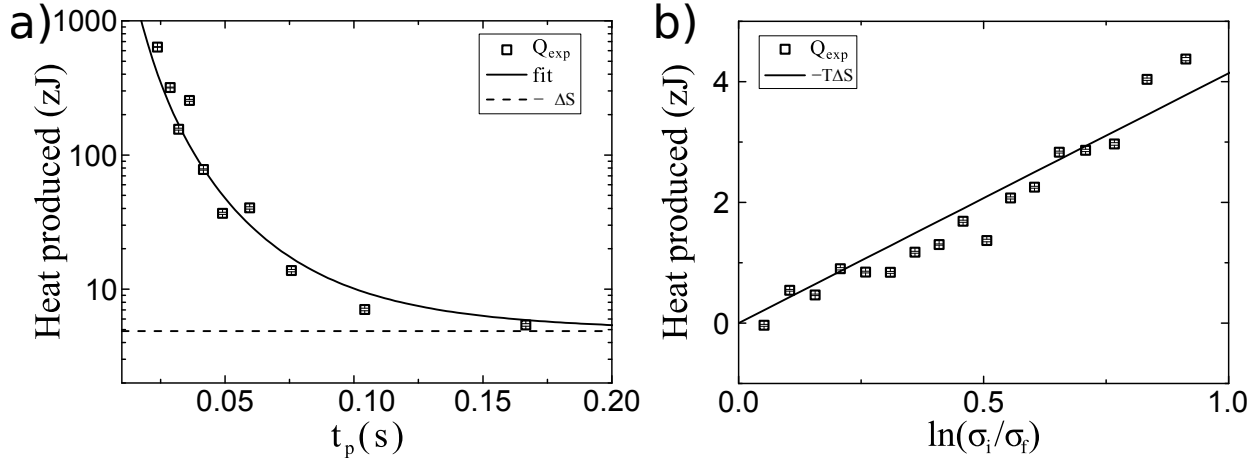}
\caption{Experimental results of produced heat. (a) Produced heat for a single refresh as function $t_p$. By increasing $t_p$ the produced heat tends to the lower bound $Q=-T\Delta S$. Squares represent the heat from the experiment, while the solid line is the fit with the Zener dissipative model. (b) Produced heat in the quasi-static regime during a single refresh operation for different entropy variations. Squares represent the estimated heat from experiments while the solid line is given by eq.\eqref{eq:equation3}.}\label{fig:figure_4}
\end{figure}

In Fig.\ref{fig:figure_4}.a we show the measured values of $Q$ required to perform a single refresh operation as a function of the protocol time $t_p$, for fixed $\sigma_i$ and $\sigma_f$. We can see that $Q$ approaches the minimum value given by eq.\eqref{eq:equation3} when $t_p$ increases towards the quasi-static protocol condition. This observation is confirmed for different values of $\Delta S$, as we can see from Fig. \ref{fig:figure_4}.b. There we show the measured values of $Q$ for a quasi-static protocol as a function of $-\ln(\sigma_i/\sigma_f)$.
Experimental points are given as black squares while the black solid line is the theoretical prediction from eq.\eqref{eq:equation3}. 
As it is well apparent, the minimum energetic cost, represented by the thermodynamics bound $-T\Delta S$ can be reached in the quasi-static condition.
The dissipative model behind the power law fit in Fig.\ref{fig:figure_4}.a is obtained by the Zener theory\cite{lopez,zener,zener2,zener3,saulson}, assuming that the dissipative processes can be expressed as the result of frictional forces that represent the imaginary component of a complex elastic force $−k(1+i\phi)$. In general, $\phi$ is a function of the frequency and for small damping it can be expressed as the sum over all the dissipative contributions. In our case $\phi(\nu)=\phi_{str}+\phi_{th-el}+\phi_{vis}+\phi_{clamp}$. Here $\phi_{str}$ is the structural damping ($\phi$ is independent of the frequency $\nu$), $\phi_{th-el}$ and $\phi_{vis}$ are the thermo-elastic and viscous damping that can be assumed to be proportional to the frequency for frequencies much smaller than the cantilever characteristic frequency, and $\phi_{clamp}$ represents the clamp recoil losses ($\phi(\nu) \propto \nu^3$).

Based on this result we are now in position to express the minimum fundamental cost  $Q_m$ for preserving a memory over a time $\overline{t}$ with a failure probability equal to $P_E$ as
\begin{equation}\label{eq:equation4}
Q_m=-N T\Delta S=\frac{\overline{t}}{t_R}  k_B T \ln\left(\tfrac{\sqrt{\sigma_w^2+e^{-\frac{t_R}{\tau_w}} (\sigma_i^2-\sigma_w^2)}}{\sigma_i} \right)
\end{equation}
\begin{figure}
\includegraphics[width=\linewidth]{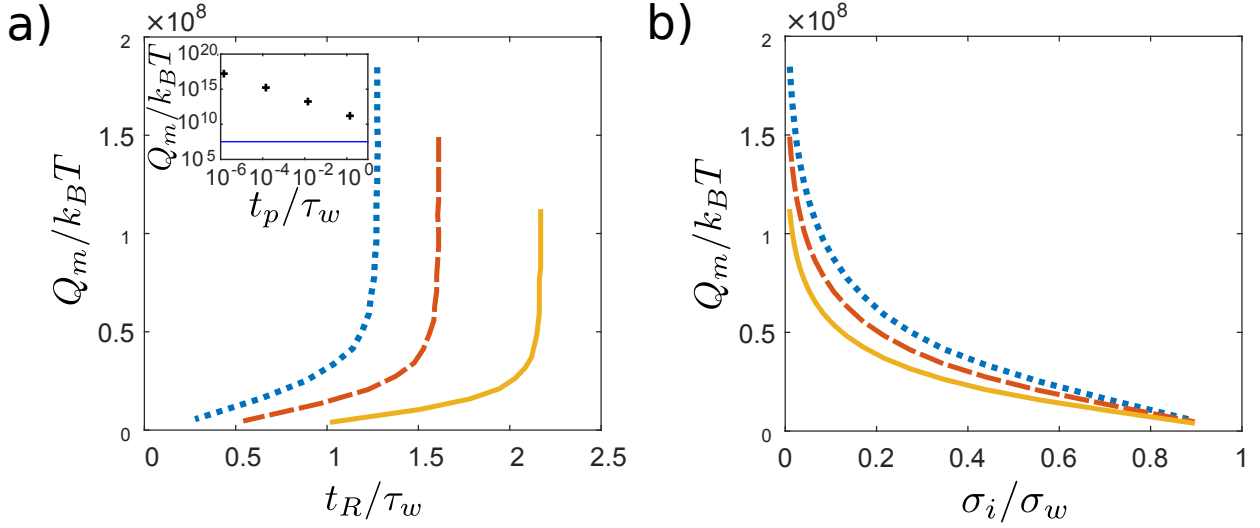}
\caption{Plots of $Q_m$ to preserve the memory for $\overline{t}=\num{1e3}\tau_k$ as a function of $t_R$ (a) and $\sigma_i$ (b). Blue (dotted) lines are obtained with $P_E=\num{1e-6}$, red (dashed) lines with $P_E=\num{1e-4}$, and yellow (solid) lines with $P_E=\num{1e-2}$. Inset in panel (a) shows the values for $Q_m$ vs $t_p/\tau_w$, when $\overline{t}=\num{1e3}\tau_k$, $P_E=\num{1e-6}$ and $t_R=\tau_w.$ A finite protocol time $t_p$, which is typical of experiments, adds an excess dissipated heat to the blue line that marks the minimum value given eq.\eqref{eq:equation4}.}\label{fig:figure_5}
\end{figure}

In Fig. \ref{fig:figure_5}.a we show the minimum energy $Q_m$ as a function of $t_R$ for a given choice of $P_E$ and $\overline{t}$. It is interesting to observe that this is an increasing function of $t_R$. In particular $Q_m$ approaches the value 0 when $t_R$ goes to 0. This indicates that it is possible, at least in principle, to preserve the memory for a time $\overline{t}$ with failure probability $P_E$ while spending zero energy. This is obtained when $t_R$ approaches 0 but it also implies that the memory is always under refresh and never available for use. Moreover $Q_m$ diverges when $t_R$ approaches a limit value $t_{RMax}$ that depends on $P_E$. In fact when $t_R \geq t_{RMax}$ the imposed conditions on $P_E$ and $\overline{t}$ cannot be satisfied. On approaching such a value, $\sigma_i$ has to become smaller and smaller, thus requiring a larger and larger energy. This is apparent in Fig. \ref{fig:figure_5}.b where we show the minimum energy $Q_m$ as function of $\sigma_i$ for a given choice of $P_E$ and $\overline{t}$. 
There, $Q_m$ goes to 0 when $\sigma_i$ goes to $\sigma_w$. This implies that it is indeed possible to preserve the memory for a time $\overline{t}$ with probability $P_E$ by spending zero energy and this is realised when we operate extremely close to the equilibrium configuration inside one well ($\sigma_i\to \sigma_w$). On the contrary, as we anticipated, $Q_m$ grows toward infinity when $\sigma_i \to 0$. 
Nonetheless, this last condition is limited by the Heisenberg uncertainty relation. By taking $\sigma_i$ to coincide the uncertainty on the position, we have $\sigma_i \geq\hbar/(2\sigma_p)$, where $\sigma_p$ is the uncertainty on the momentum. This latter quantity, for a system at thermal equilibrium, can be estimated with the equipartition theorem if we have that $k_BT$ is much greater than the energy separation between the system quantum levels. This is what happens in macroscopic devices that works at room temperature. Since the equipartition gives a finite value for $\sigma_p$, the uncertainty principle then sets a maximum accuracy on the position. This means that, for a given system, the probability distribution of the relevant degree of freedom cannot be shrunk arbitrary (see Appendix \ref{a:C}). Since $Q_m$ in Fig. \ref{fig:figure_5}.b is a monotone function we have that $Q_m$ reaches a finite maximum value for the minimum allowed $\sigma_i$. 

The existence of a minimum $\sigma_i$ has a more important consequence: it sets a limit on our capability to preserve a given memory forever. This is apparent when we use eq.\eqref{eq:equation1} to explicitly write
\begin{equation}\label{eq:equation5}
\overline{t} = t_R \ln(1-P_E)/\ln(1-P_0).
\end{equation}
Once we set $P_E$ and select a finite $t_R$, we can make $\overline{t}$ as large as we want by properly selecting $P_0$ small enough. However, the existence of a finite minimum $\sigma_{i}$ implies that $P_0$  can never be smaller than a nonzero minimum value, thus $\overline{t}$ reaches a finite maximum at best.
To estimate such a maximum $\overline{t}$ in practical memories, we consider a micromechanical memory device like the one in Ref. \citenum{neri}. If we assume the distance between the two wells $x_m = \SI{1e-9}{\metre} $ and a refresh period $t_R= \SI{6.6e-3}{\second}$, we have that the minimum $\sigma_{i}= \SI{9.6e-20}{\metre}$. If we set $P_E =\num{1e-6}$ then the maximum value for $\overline{t}$ is approximately 2 years. On the other hand, if we set $P_E =\num{1e-4}$ then the maximum time $\overline{t}$ is approximately $200$ years.

Finally we briefly discuss the role of the protocol time $t_p$. As we have seen above from the experiment, the minimum fundamental bound $Q_m$ can be reached only in the quasi-static regime where $t_p$ is non negligible. 
This condition sets a minimum value for $t_R$, such that $t_R \geq t_p$ and prevents the possibility to perform the experiment at zero energy expenditure.
Moreover for any finite $t_p$ frictional losses add to the minimum refresh cost $Q_m$, as it is clearly visible from the experimental data in Fig. \ref{fig:figure_4}.a. 

In order to identify a general estimate of the overall energy cost with a finite $t_p$, for a given choice of $P_E$, $\overline{t}$ and $t_R$, we use the formal tools developed in Refs. \citenum{espositobroeck}, \citenum{chiuchiudiamantini}, \citenum{ciliberto_ESE} to obtain a final condition of the protocol with the desired value of $\sigma_i$. The results are shown in the inset of Fig. \ref{fig:figure_5}.a. There we see that the dissipated energy $Q_m$ is an inverse function of $t_p$, and that finite protocol times increase the energetic cost to refresh one bit by orders of magnitude respect to the minimum cost prescribed by eq.\eqref{eq:equation4}.

In conclusion we studied the energy cost associated with memory preservation. We have introduced a physical model for the refresh procedure and realised an experiment in order to measure the amount of work performed during the refresh operation. Our study indicates that, in principle, we can preserve a digital memory for a given finite time with a given error probability while spending an arbitrarily little amount of energy. This is accomplished with refresh procedures that are performed arbitrarily often (Fig. \ref{fig:figure_5}.a) and/or arbitrarily close to thermal equilibrium (Fig. \ref{fig:figure_5}.b). In practical cases however the existence of frictional forces introduces a lower limit on the refresh interval $t_R \geq t_p$ and this imply a non-zero minimum energy expenditure (Fig. \ref{fig:figure_5}.a inset).
We have also shown that, by the moment that the Heisenberg uncertainty principle implies the existence of a minimum width for the initial probability density of the memory device, any refresh strategy will inevitably fail after a finite time. 

\begin{acknowledgments}
The authors gratefully acknowledge financial support from the European Commission (H2020, Grant agreement no: 732631, OPRECOMP, FPVII, Grant agreement no: 318287, LANDAUER and Grant agreement no: 611004, ICT- Energy) and ONRG grant N00014-11-1-0695.
\end{acknowledgments}

\appendix
\section{Computation of Fig.\ref{fig:figure_2}}\label{a:A}
To compute Fig.\ref{fig:figure_2}, we take
\begin{equation}
p(x,0)=\frac{\exp\left(-\frac{(x-1)^2}{2\sigma_i^2}\right)}{\sqrt{2\pi}\sigma_i}
\end{equation}
as initial condition for eq.\eqref{eq:FP}. In particular, $\sigma_i$ is such that $p(x,t)$ broadens inside the right well of $U(x)$ before developing a clean-cut second peak in the left well of the potential. We then solve eq.\eqref{eq:FP} with the Matlab \texttt{pdepe} function. With the solution we compute
\begin{equation}
P_0(t_R)=\int_{-\infty}^{0} p(x,t_R)
\end{equation}
for different refresh times $t_R$, and then we evaluate the failure probability
\begin{equation}
P_E=1-\left(1-P_0(t_R)\right)^\frac{t}{t_R}
\end{equation}
for different values of $t_R$ and $t\gg t_R$. As a last step, we use a spline fit of $P_E$ to sample $t_R$ for different values of $P_E$ and $t$. The results obtained in this way are plotted in Figure 2. These results are obtained with $T = 1/8$ which corresponds to $\tau_k = \num{5.3e4}\tau_w$. 
 
\section{Derivation of eq.\eqref{eq:equation3} and eq.\eqref{eq:equation2}}\label{a:B}
We derive here two important equations given in the main text, namely eq.\eqref{eq:equation3} and eq.\eqref{eq:equation2}.
We start with eq.\eqref{eq:equation2}. To derive it we assume that $T \ll 1$. This simplifies the mathematical description of the system as it implies that the intra-well relaxation mechanisms of the system are faster than the inter-well ones. If we are interested in intra-well mechanism only, then a satisfactory form for the dimensionless $p(x,t)$ is
\begin{subequations}\label{eq:twopeaks}
\begin{align}
p(x,t)=&p_0(x,t)+p_1(x,t)\\
p_0(x,t)=&P_0\frac{\exp\left(-\frac{(x+1)^2}{2\sigma(t)^2}\right)}{\sqrt{2\pi}\sigma(t)}\\
p_1(x,t)=&(1-P_0)\frac{\exp\left(-\frac{(x-1)^2}{2\sigma(t)^2}\right)}{\sqrt{2\pi}\sigma(t)}\\
\end{align}
\end{subequations}
where $P_0$ is, to all effects, constant over time. We substitute eq.\eqref{eq:twopeaks} in eq.\eqref{eq:FP} and then we approximate eq.\eqref{eq:duffing} with an harmonic potential by Taylor-expanding around $x=\pm 1$. This yields two distinct equations
\begin{subequations}\label{eq:doppia_FP}
\begin{align}
\pdv{p_0}{t}-8\left(p_0+(x+1)\pdv{p_0}{x}\right)-T\pdv{^2p_0}{x^2}&=0\\
\pdv{p_1}{t}-8\left(p_1+(x-1)\pdv{p_1}{x}\right)-T\pdv{^2p_1}{x^2}&=0\\
\end{align}
\end{subequations}
where used the fact that $p_1(x,t)$ ($p_0(x,t)$) can't affect the dynamics of the system in the left (right) well of $U(x)$ if $\Delta U \ll k_BT$. Eq.\eqref{eq:doppia_FP} can be combined into
\begin{equation}
\begin{aligned}
&\int_{-\infty}^{\infty}\left(\pdv{p_0}{t}-8\left(p_0+(x+1)\pdv{p_0}{x}\right)-T\pdv{^2p_0}{x^2} \right)(x+1)^2 \mathrm{d}x\\
&+\int_{-\infty}^{\infty} \left(\pdv{p_1}{t}-8\left(p_1+(x-1)\pdv{p_1}{x}\right)-T\pdv{^2p_1}{x^2}\right)(x-1)^2 \mathrm{d}x\\
&=0,
\end{aligned}
\end{equation}
which reduces to
\begin{equation}\label{eq:sigma_eq}
\pdv{\sigma(t)^2}{t}+16\sigma(t)^2-2T=0.
\end{equation}
Eq.\eqref{eq:sigma_eq} describes the time evolution of $\sigma(t)$ when intra-well relaxation mechanisms occurs. Its analytic solution for an initial condition $\sigma(0)=\sigma_i$ is
\begin{equation}\label{eq:sigma_t}
\sigma(t)=\sqrt{\frac{T}{8}+\exp\left(-16 t\right)\left(\sigma_i^2- \frac{T}{8}\right)}
\end{equation}
which is the dimensionless version of the eq.\eqref{eq:equation2} given the main text.

To compute eq.\eqref{eq:equation3} we recollect that we defined the ``refresh operation restores a non-equilibrium condition by shrinking the width of each peak of $p(x,t)$.'' without error corrections. If we assume that the refresh protocol preserves the symmetry of $U(x)$, then $p(x,t)$ can be written as eq.\eqref{eq:twopeaks} during the whole refresh  procedure. As a consequence, the sole effect of a refresh operation with duration $t_p$ is to transform
\begin{equation}\label{eq:init_cond}
p(x,t)=P_0(t)\frac{\exp\left(-\frac{(x+1)^2}{2\sigma(t)^2}\right)}{\sqrt{2\pi}\sigma(t)}+(1-P_0(t))\frac{\exp\left(-\frac{(x-1)^2}{2\sigma(t)^2}\right)}{\sqrt{2\pi}\sigma(t)}\\
\end{equation}
into
\begin{equation}\label{eq:finit_cond}
p(x,t+t_p)=P_0(t)\frac{\exp\left(-\frac{(x+1)^2}{2\sigma_i^2}\right)}{\sqrt{2\pi}\sigma_i}+(1-P_0(t))\frac{\exp\left(-\frac{(x-1)^2}{2\sigma_i^2}\right)}{\sqrt{2\pi}\sigma_i}\\
\end{equation}
where $\sigma(t)$ is given by eq.\eqref{eq:sigma_t}, $P_0(t)$ is fitted from the numerical solution of eq.\eqref{eq:FP} with eq.\eqref{eq:duffing}, and $\sigma_i=\sigma(0)$.
We now use the Gibbs entropy definition
\begin{equation}
S(t)=-k_B\int_{-\infty}^{\infty} p(x,t)\ln{p(x,t)} \mathrm{d} x
\end{equation}
to compute the entropy variation $\Delta S=S(t+t_p)-S(t)$ of the refresh protocol. Because of the $\Delta U\gg k_BT$ assumption, we have that
\begin{equation}
\begin{aligned}
&\Delta S\approx -k_B\Bigl(\int_{-\infty}^{\infty} \frac{\mathrm{e}^{-\frac{x^2}{2\sigma_i^2}}}{\sqrt{2\pi}\sigma_i}\ln\Bigl(\frac{\mathrm{e}^{-\frac{x^2}{2\sigma_i^2}}}{\sqrt{2\pi}\sigma_i} \Bigr) \mathrm{d}x \\
&-\int_{-\infty}^{\infty} \frac{\mathrm{e}^{-\frac{x^2}{2\sigma(t)^2}}}{\sqrt{2\pi}\sigma(t)}\ln\Bigl(\frac{\mathrm{e}^{-\frac{x^2}{2\sigma(t)^2}}}{\sqrt{2\pi}\sigma(t)} \Bigr) \mathrm{d}x\Bigr),
\end{aligned}
\end{equation}
which reduces to 
\begin{equation}\label{eq:DeltaS}
\Delta S\approx k_B\ln\left(\frac{\sigma_i}{\sigma(t)}\right).
\end{equation}
By using eq.\eqref{eq:DeltaS} with $t=t_R$ we obtain the eq.\eqref{eq:equation3} presented in the main text.

\section{Minimum value for $\sigma_i$}\label{a:C}
We discuss here the existence of the minimum possible value for $\sigma_i$. First of all, we observe that $\sigma_i\to0$ is a singular limit in eq.\eqref{eq:DeltaS}. This is inconsistent with the third law of thermodynamics, so there must be a minimum value for $\sigma_i$. This is given by the Heisenberg uncertainty principle. In the best case scenario this reads
\begin{equation}
\sigma_x\sigma_p=\frac{\hbar}{2}.
\end{equation}
where $\sigma_x$ ($\sigma_p$) is the uncertainty on the position $x$ (momentum $p$). According to the equipartition theorem,
\begin{equation}
\sigma_p=m{\sqrt{\langle v^2\rangle-\langle v\rangle^2}}={\sqrt{mk_B T}},
\end{equation}
so we have that
\begin{equation}\label{eq:sigma_x}
\sigma_x=\frac{\hbar}{2\sqrt{mk_BT}}
\end{equation}
Eq.\eqref{eq:sigma_x} sets the minimum possible uncertainty for $\sigma_x$. Since $\sigma_i$ describes the uncertainty of the initial $x$ value, we therefore have that $\sigma_i\geq \sigma_{iMin} =\frac{\hbar}{2\sqrt{mk_BT}}$. The existence of a $\sigma_{iMin}$ implies that, even at $t=0$, the probability of error $P_0$ is greater than zero. Clearly, this does not exclude that one can have a smaller $\sigma_i$ by accepting a larger $\sigma_p$. This would imply to operate the memory out of the thermal equilibrium, growing the dissipated energy well above the fundamental minimum.

\end{document}